\documentclass[journal,onecolumn]{IEEEtran}
\usepackage{cite}
\usepackage{url}

\ifCLASSINFOpdf
\else
\fi
\usepackage{amssymb}
\usepackage{amsmath}
\usepackage{multirow}

\usepackage{graphicx}
\usepackage{epstopdf}
\begin{document}
%
% paper title
% Titles are generally capitalized except for words such as a, an, and, as,
% at, but, by, for, in, nor, of, on, or, the, to and up, which are usually
% not capitalized unless they are the first or last word of the title.
% Linebreaks \\ can be used within to get better formatting as desired.
% Do not put math or special symbols in the title.
\title{Regional Total Electron Content Map Generation based on Compressive Sensing}
%
%
% author names and IEEE memberships
% note positions of commas and nonbreaking spaces ( ~ ) LaTeX will not break
% a structure at a ~ so this keeps an author's name from being broken across
% two lines.
% use \thanks{} to gain access to the first footnote area
% a separate \thanks must be used for each paragraph as LaTeX2e's \thanks
% was not built to handle multiple paragraphs
%

%\author{Michael~Shell,~\IEEEmembership{Member,~IEEE,}
%        John~Doe,~\IEEEmembership{Fellow,~OSA,}
%        and~Jane~Doe,~\IEEEmembership{Life~Fellow,~IEEE}% <-this % stops a space
%\thanks{M. Shell was with the Department
%of Electrical and Computer Engineering, Georgia Institute of Technology, Atlanta,
%GA, 30332 USA e-mail: (see http://www.michaelshell.org/contact.html).}% <-this % stops a space
%\thanks{J. Doe and J. Doe are with Anonymous University.}% <-this % stops a space
%\thanks{Manuscript received April 19, 2005; revised August 26, 2015.}}

\author{Cansu~Sunu,
	Cenk~Toker% stops a space
	%\\ Department of Electrical and Electronics Engineering, Hacettepe University, Ankara, Turkey.
	\thanks{C. Sunu and C. Toker are with the Department
		of Electrical and Electronics Engineering, Hacettepe University, Ankara, Turkey.}}

\maketitle

% As a general rule, do not put math, special symbols or citations
% in the abstract or keywords.
\begin{abstract}
Ionosphere has an important role in long distance HF communications, satellite communications and global navigation systems. Ionosphere is a plasma medium which arises due to solar and cosmic radiation, and the amount of ionization is highly time and location dependent. Eventually, the current state of the ionosphere should be continuously monitored with high accuracy.
Total Electron Content (TEC) maps are being used to investigate the state of the ionosphere. There are online services which provide TEC maps, however they mostly have low spatial and temporal resolution and the techniques used for generating these maps are generally not accessible. 

TEC maps can also be generated from GNSS/GPS based Continuously Operating Receiver Stations (CORS) network measurements. Unfortunately, the GNSS/GPS receiver networks are not dense enough to form a map directly. Therefore, an algorithm should be used to estimate the TEC values at coordinates without a receiver.
Based on the observation that TEC maps possess a high degree of sparsity, we propose a modified compressive sensing technique for generating regional TEC maps by using the sparse dataset obtained from a CORS network. 

We evaluate the performance of the proposed technique both over synthetically generated TEC maps which mimic the common characteristics of the ionosphere, and also over actual measurements taken over the Turkish National Permanent GPS Network (TNPGN) Active. Our analysis reveals that the proposed technique can produce TEC maps with high accuracy and resolution. We also demonstrate the superiority of our technique over other TEC map generation techniques found in the literature.
%Kriging estimations show spurious chequered like patterns, but CS presents no such patterns. 
\end{abstract}
% Note that keywords are not normally used for peerreview papers.
% For peer review papers, you can put extra information on the cover
% page as needed:
% \ifCLASSOPTIONpeerreview
% \begin{center} \bfseries EDICS Category: 3-BBND \end{center}
% \fi
%
% For peerreview papers, this IEEEtran command inserts a page break and
% creates the second title. It will be ignored for other modes.
\IEEEpeerreviewmaketitle
\vspace*{-0.3cm}
\section{Introduction}
% The very first letter is a 2 line initial drop letter followed
% by the rest of the first word in caps.
% 
% form to use if the first word consists of a single letter:
% \IEEEPARstart{A}{demo} file is ....
% 
% form to use if you need the single drop letter followed by
% normal text (unknown if ever used by the IEEE):
% \IEEEPARstart{A}{}demo file is ....
% 
% Some journals put the first two words in caps:
% \IEEEPARstart{T}{his demo} file is ....
% 
% Here we have the typical use of a "T" for an initial drop letter
% and "HIS" in caps to complete the first word.
\IEEEPARstart{I}{onosphere}, located at an altitude of between 90 km and 1100 km above the Earth, is a plasma medium ionized by high energy solar and cosmic radiations. Ionosphere plays an important role in long distance communications, satellite communications and global navigation systems since all these signals pass through the ionosphere as the propagation medium.

Space Weather (SW) phenomena such as solar flares, and ionospheric storms strongly affect the structure of the ionosphere. The performance of the systems mentioned above can significantly be deteriorated and even sometimes they cannot be used temporally due to ionospheric disturbances. The ionosphere varies considerably with time, location, activities of sun, and geomagnetic storms, hence the structural variations of the ionosphere have to be observed and tracked continuously \cite{CToker}. The key parameter used to express the ionospheric structure is the Total Electron Content (TEC). TEC, which is defined as the total number of electrons contained in a hypothetical semi-infinite cylinder with a base area of $1$ m$^2$ positioned on the ground and aligned in the vertical direction. Unit of TEC is TECU and $1$ TECU is $10^{16}$ electrons/m$^2$.

TEC is not directly measurable, but its value is estimated by indirect measurements made using ground- or space-based measurement systems. Ground-based systems are ionosondes, backscatter radars, and noncoherent backscatter radars etc. Space-based systems depend on systems such as Global Positioning System (GPS), GLONASS, and TOPEX/Poseidon. Since GPS is a global system and there are several Continuously Operating Receiver Stations (CORS) networks composed of hundreds of GPS receivers deployed both globally and also regionally such as International GNSS Stations (IGS), Regional Reference Frame Sub-Commission for Europe (EUREF), Turkish National Permanent GPS Network (TNPGN) Active, etc., TEC measurements from GPS signals is a widely used technique, \cite{GPS1,GPS2,GPS3,GPS4}.

Temporal and spatial variability of the ionosphere can be represented by the temporal and spatial correlation of the electron density of the ionosphere. Quiet and disturbed days are investigated in \cite{ISayin}, and it is shown that temporal correlation varies from $3$ to $25$ mins for quiet days and from $3$ to $15$ mins for disturbed days \cite{Yenen}. Shorter correlation time and correlation distance values are indications of high variability in the ionosphere. The spatial and temporal behaviour of the ionosphere can be effectively analysed using two dimensional TEC maps \cite{CSunu,Mannucci,Zhao,Liu,Meza}. In \cite{CSunu}, global $1^\circ$ (lat) $\times$ $1^\circ$ (lon) TEC maps are analysed using 2-Dimensional Discrete Cosine Transform (2D-DCT), where it is shown that the maps can be represented by using a small number of 2D-DCT coefficients. Specifically, it was shown that $90\%$ of the energy content of the DCT coefficients of a TEC map is contained within the largest $20$ coefficients of the transform of a $180\times360=64800$ pixel TEC map. This observation demonstrates the high sparsity level of TEC maps in the DCT domain. 
%The approximate maps of different degrees of sparsity have more than $90$ $\%$ of the energy of the original map. 
Similar results on the sparsity level of the TEC maps for the mid latitude region are also shown in \cite{CToker}. 

Compressive Sensing (CS) is a method successfully applied to signals which are highly sparse in a certain transform domain. In contrast to the conventional sampling theorem, where uniformly spaced samples of the signal are required with a sampling frequency of at least twice the bandwidth of the signal of concern, a much smaller number of samples with non-uniform (possibly random) spacing are adequate to recover the original signal from these sparse samples in CS, given that the signal is highly sparse in a transform domain.
%Compressive sensing (CS) is a technique that recovering a signal, which shows sparsity in nature or a transform domain, using a set of samples taken under the Nyquist rate of twice the highest frequency present in the signal of interest. In compressive sensing, the set of samples is not necessarily be uniformly spaced as in conventional sampling. 
Signal recovery is done by using optimization problems or iterative algorithms \cite{CS_TheAndApp,Candes,CSAlgo}. The concept and applications of compressive sensing are reviewed in \cite{Rani} in detail. Some of the major application areas of CS are radar imaging systems, biomedical applications, pattern recognition, and communications networks \cite{Rani71,Rani72,Rani75,CS65,CS178}. In this paper, the compressive sensing technique is applied to the regional TEC map estimation based on the sparse nature of the TEC maps. Application of compressive sensing to the TEC map generation presented in this paper is a novel technique. 

TEC maps are being published online by several research/government  organizations through their webpages \cite{NASAJPL,DLR,NOAA,SWS,CODE}. However these maps have low spatial and temporal resolution (e.g. NASA-JPL (IARS) \cite{NASAJPL} produces $2.5^\circ$ (lat) $\times$ $5^\circ$ (lon) maps at each $5$ minutes, DLR (IMPC) \cite{DLR} produces $2.5^\circ$ (lat) $\times$ $5^\circ$ (lon) global and $2^\circ$ (lat) $\times$ $2^\circ$ (lon) regional (Europe) maps at an update rate of 15 minutes, SWS \cite{SWS} produces global/regional maps at each $15$ minutes, UNIBE (CODE-GIM) \cite{CODE} maps are produced at a resolution of $2.5^\circ$ (lat) $\times$ $5^\circ$ (lon) and $2$ hours, where the near real-time maps are produced through a ``rapid" process and final maps are provided a couple of days later. Moreover, the processes being used to produce these maps are not open to the public access.

\begin{figure}[!t]
	\centering
	\includegraphics[width=\columnwidth]{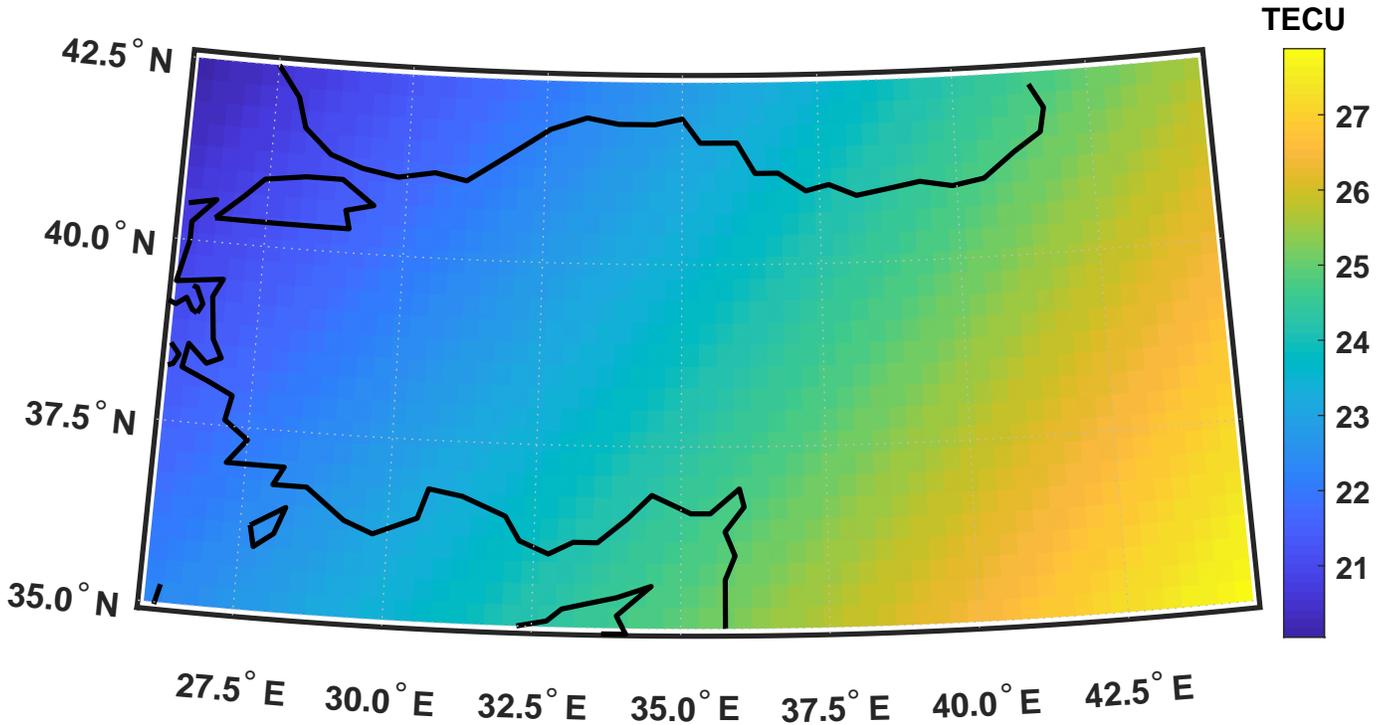}
	\caption{Synthetic map 1: Linear trend. Represents the dawn of a quite day.}
	\label{fig:SS2MAP}
\end{figure}
\setlength{\textfloatsep}{0.2\baselineskip plus 0.2\baselineskip minus 0.5\baselineskip}
\begin{figure}[!t]
	\centering
	\includegraphics[width=\columnwidth]{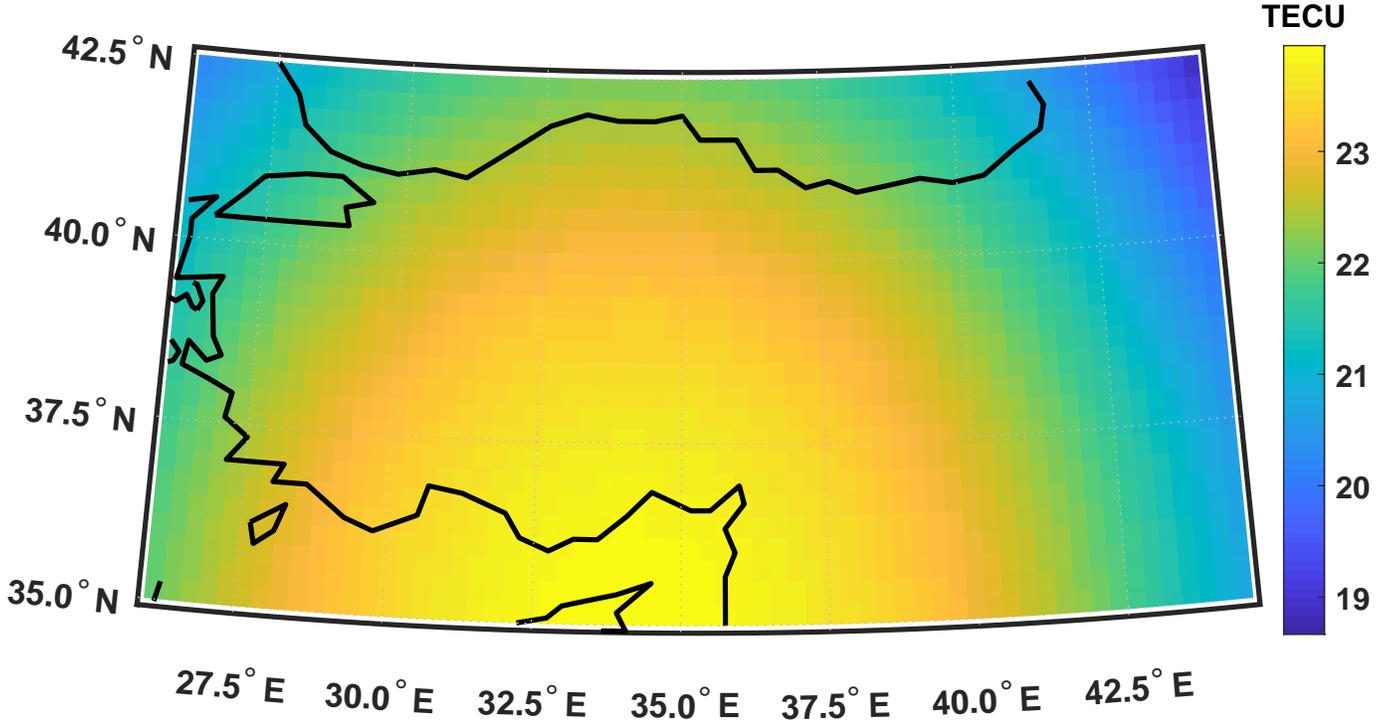}
	\caption{Synthetic map 2: Quadratic trend. Represents the midday of a quiet day.}
	\label{fig:SS3MAP}	
\end{figure}

Eventually, for the purpose of real-time monitoring of the ionosphere, observing ionospheric anomalies such as travelling ionospheric disturbances (TIDs), and making real-time corrections for GNSS based navigation systems, the services mentioned above may not be adequate. In this paper, we propose a technique which produces TEC maps with a spatial resolution better than $0.5^\circ$ (lat) $\times$ $0.5^\circ$ (lon) and a temporal resolution of as low as $30$ sec.s. Moreover, TEC maps can be generated as soon as the GNSS/GPS data is available, making our technique ``real time".

An alternative to our proposed technique is based on Kriging \cite{Sayin2008,Deviren2018}. In this technique, TEC measurements from GNSS receivers are used to construct the TEC map, however, the reconstruction performance deteriorates with increasing resolution, and artifacts  appear, especially in regions where extrapolation needs to be done. Our technique is superior to Kriging both in the sense of improved resolution and also reconstruction performance, as will be compared in this paper.

Section II introduces the system model, and discusses 2D-DCT domain representation of a TEC map. Sparsity levels of synthetically generated maps, and the optimization problem which is used to recover the map from a set of samples are also provided in Section II. In Section III, the performance of the proposed technique on synthetically generated maps in the sense of normalized error energy is analysed. In Section IV, map estimations are made using GPS-TEC data from Turkish National Permanent GPS Network (TNPGN) Active and results are compared with the Kriging method \cite{Sayin2008,Deviren2018}. Finally, Section V concludes the paper.
\vspace*{-0.2cm}
\section{System Model and Problem Formulation}
In this section, we construct the mathematical basis of the problem and provide an optimization problem formulation for the reconstruction of the TEC maps.
\vspace*{-0.4cm}
\subsection{Notation}
Let a TEC map of a ``rectangular" region between coordinates $\phi_\text{min}$ and $\phi_\text{max}$ in latitude and $\lambda_\text{min}$ and $\lambda_\text{max}$ in longitude (following the geodetic coordinate system convention) at a given time be represented by a $P\times Q$ matrix $\mathbf{U}$
%The TEC map at a given time with $P$ and $Q$ points in lattitude and longitude, respectively, can be represented in the matrix form such as
\begin{equation} \label{Eq:MapMatrix}
\mathbf{U} = \begin{bmatrix}
U_{0,0} & U_{0,1} & \cdots & U_{0,Q-1} \\
U_{1,0} & U_{1,1} & \cdots & U_{1,Q-1} \\
\vdots & \vdots & \ddots & \vdots \\
U_{P-1,0} & U_{P-1,1} & \cdots & U_{P-1,Q-1}
\end{bmatrix}.
\end{equation}
If a grid structure on the map is visualized with resolution $\Delta\phi$ in latitude and $\Delta\lambda$ in longitude, the real-valued scalar $U_{p,q}$ gives the TEC value at the coordinate ($\phi_\text{min}+p\Delta\phi$, $\lambda_\text{min}+q\Delta\lambda$). We assume that the lower-left pixel of the map is the origin.

\begin{figure}[!t]
	\centering
	\includegraphics[width=\columnwidth]{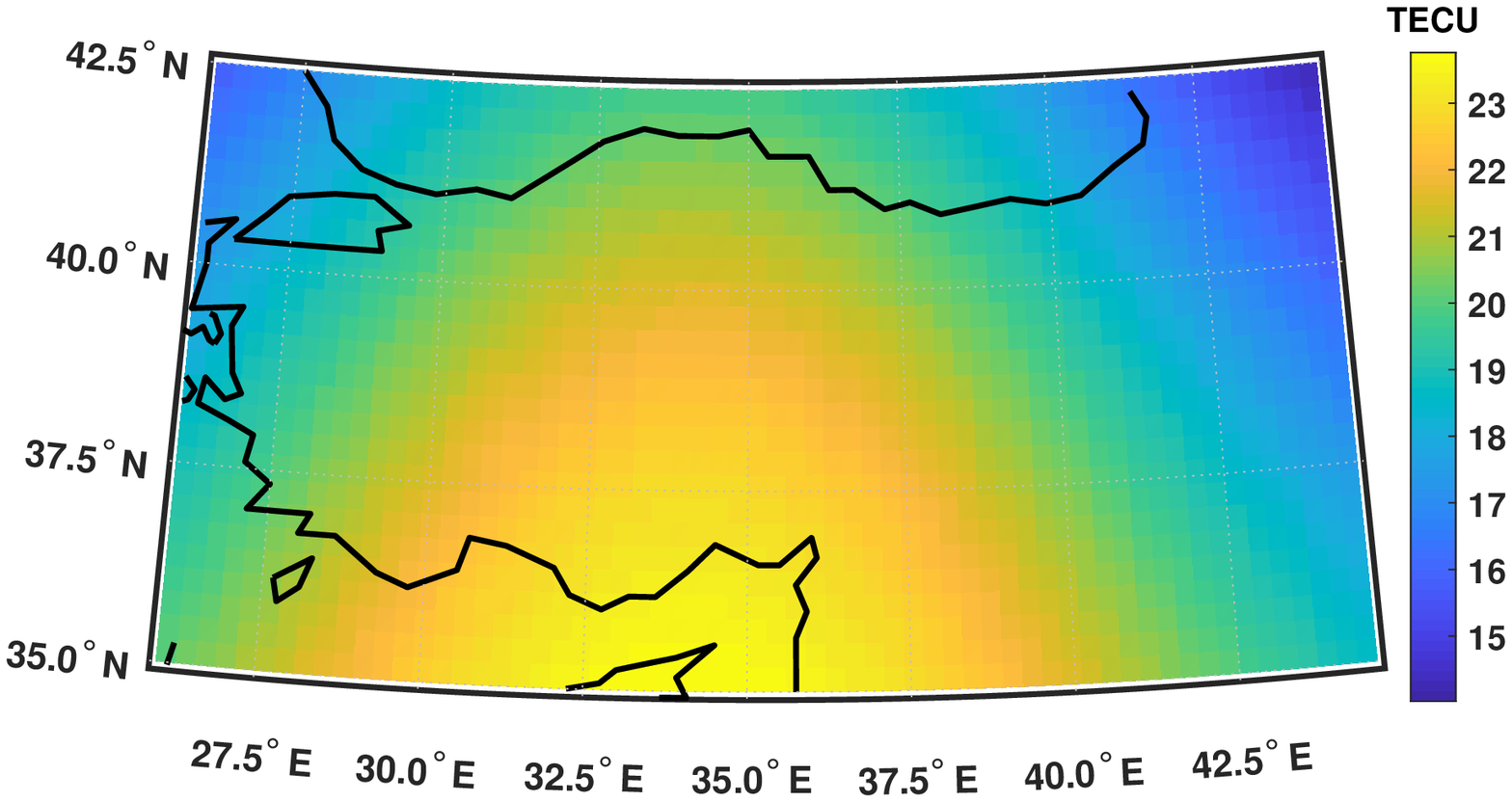}
	\caption{Synthetic map 3: Gaussian trend. Represents the midday of a quiet day.}
	\label{fig:SS4MAP}
\end{figure}
\begin{figure}[!t]
	\centering
	\includegraphics[width=\columnwidth]{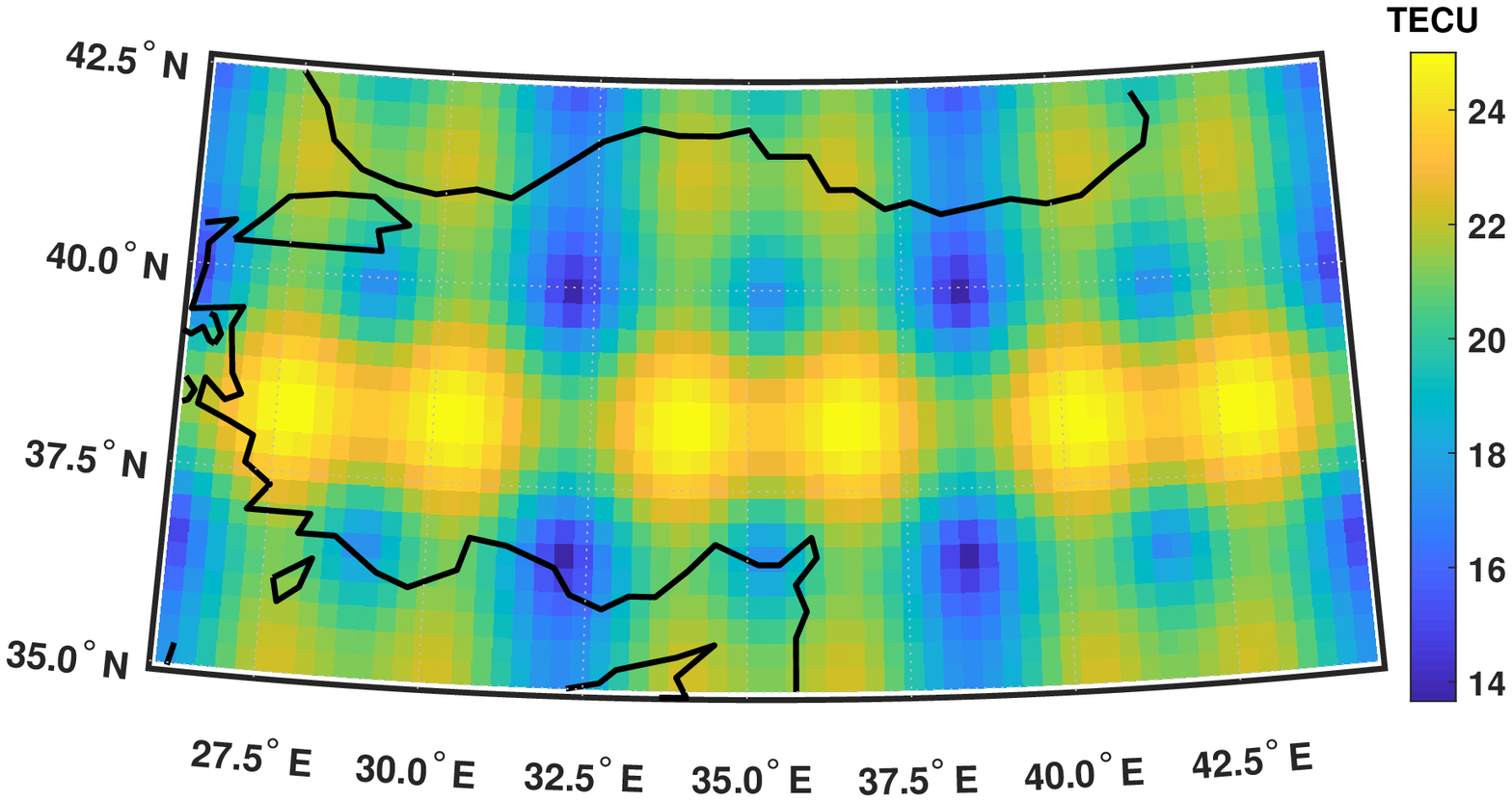}
	\caption{Synthetic map 4: Mixture of Gaussian trend and sinusoidal disturbance. Represents the midday of a disturbed day. }
	\label{fig:SS6MAP}
\end{figure}

Let the orthonormal bases for 2D-DCT be defined as
\small
\begin{equation} \label{Eq:DCTbasis}
D_{k,l}(p,q) = \alpha_k \alpha_l \cos\Bigg(\frac{\pi(2p+1)k}{2P}\Bigg) \cos\Bigg(\frac{\pi(2q+1)l}{2Q}\Bigg)
\end{equation}
\normalsize
where $\alpha_k$ and $\alpha_l$ are  normalization factors and defined as $\alpha_k=\sqrt{\frac{1}{P}}$, if $k=0$ and $\alpha_k=\sqrt{\frac{2}{P}}$, if $k=1,...,P-1$, and $\alpha_l=\sqrt{\frac{1}{Q}}$, if $l=0$ and $\alpha_l=\sqrt{\frac{2}{QP}}$, if $l=1,...,Q-1$, respectively. 

Using (\ref{Eq:DCTbasis}),  each entry of the matrix $\mathbf{U}$ in (\ref{Eq:MapMatrix}) can be represented in terms of the 2D-DCT coefficients, $S_{k,l}$, as 
\begin{equation} \label{Eq:DctDecomp}
U_{p,q} = \sum_{k=0}^{P-1} \sum_{l=0}^{Q-1} S_{k,l} D_{k,l}(p,q).
\end{equation}

The $N\times 1$ map vector is defined as
\begin{eqnarray} \label{eq:u_vec}
\mathbf{u} &=& \text{vec}(\mathbf{U}) \nonumber \\
&=& \begin{bmatrix}
u_{0} \quad u_{1} \quad u_{2} \cdots \quad u_{N-1}
\end{bmatrix}^T
\end{eqnarray}
where $u_n = U_{p,q}$ with $n = Pq+p$ and $N = PQ$. Similarly, the $N\times 1$ 2D-DCT coefficients vector is defines as
\begin{equation} \label{Eq:DCTcoeff}
\mathbf{s} = \begin{bmatrix}
s_{0} \quad s_{1} \quad s_{2} \cdots \quad s_{N-1}
\end{bmatrix}^T
\end{equation}
where $s_t = S_{k,l}$ with $t = Pl+k$.  Using (\ref{Eq:DctDecomp})-(\ref{Eq:DCTcoeff}), we can write
\begin{equation} \label{Eq:Transform}
\mathbf{u} = \mathbf{D} \mathbf{s}
\end{equation}
where $\mathbf{D}$ is an $N\times N$ transform matrix and given by
%\small
%\begin{eqnarray}
%\mathbf{D} = \begin{bmatrix}
%D_{0,0}(0,0) &  \cdots & D_{P-1,Q-1}(0,0) \\
%D_{0,0}(1,0) & \cdots & D_{P-1,Q-1}(1,0) \\
%\vdots &  \vdots & \vdots \\
%D_{0,0}(P-1,Q-1) & \cdots & D_{P-1,Q-1}(P-1,Q-1)
%\end{bmatrix}.
%\end{eqnarray}
%\normalsize

\begin{figure}[!t]
	\centering
	\includegraphics[width=\columnwidth]{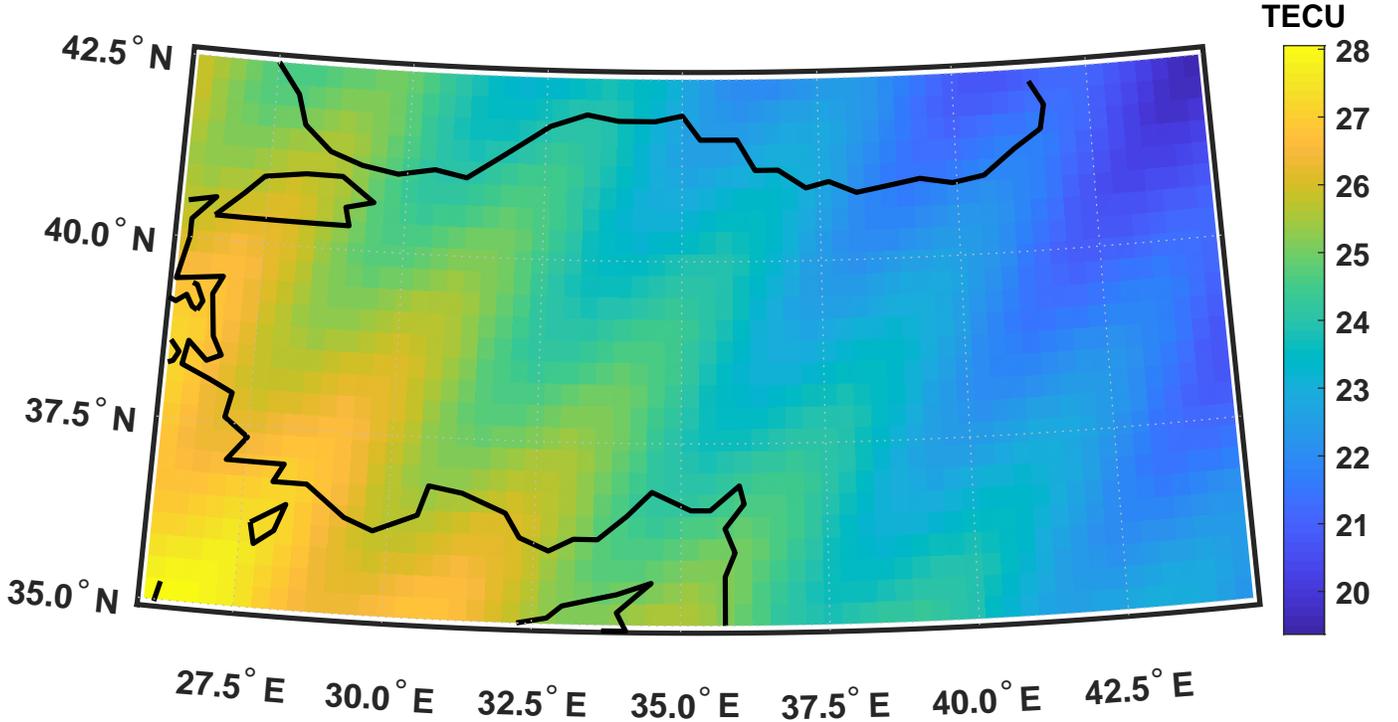}
	\caption{Synthetic map 5: Mixture of linear trend and sinusoidal disturbance. Represents the dusk of a disturbed day.}
	\label{fig:SS7MAP}
\end{figure}
\begin{equation}
\footnotesize
\mathbf{D} = \begin{bmatrix}
\tilde{D}_{0}(0)  &  \cdots  & \tilde{D}_{t}(0)  &  \cdots & \tilde{D}_{N-1}(0) \\
\tilde{D}_{0}(1)  &  \cdots  & \tilde{D}_{t}(1)  &  \cdots & \tilde{D}_{N-1}(1) \\
\vdots &  \vdots & \vdots &  \vdots & \vdots  \\
\tilde{D}_{0}(N-1)  &  \cdots  & \tilde{D}_{t}(N-1)  &  \cdots & \tilde{D}_{N-1}(N-1)
\end{bmatrix}
\normalsize
\end{equation}
where $\tilde{D}_{t}(n) = D_{k,l}(p,q)$ with $n = Pq+p$ and  $t = Pl+k$.
\begin{table}[!t]
	\renewcommand{\arraystretch}{1.8}
	\caption{Mathematical representation of Synthetic Maps (SMs)}
	\label{fig:SS_Formulation}
	\centering
	\begin{tabular}{|c|p{35pt}l|}
		\hline
		SM 1 &  TEC$(\phi,\lambda)= $ & $25 - 0.3\phi + 0.3\lambda$\\ \hline
		SM 2 &  TEC$(\phi,\lambda)=$ & $ 25 - 0.3(\phi-34)^2 + 0.3(\lambda-35)^2$\\ \hline
		SM 3 &  TEC$(\phi,\lambda)=$ & $ 20 + 5e^{-(\frac{\phi-34}{10})^2 -(\frac{\lambda-35}{7})^2}$\\ \hline
		SM 4 &  TEC$(\phi,\lambda)=$ & $ 21 + 6 \sqrt{\cos^2(\phi-35)+\sin^2(\lambda-32)} $ \\ 
		& & $- 6 e^{0.25\{\cos(\phi-35)+\cos(\lambda-32) \}} $ \\ \hline
		SM 5 &  TEC$(\phi,\lambda)=$ & $ 46 - 0.3\phi - 0.3\lambda +0.5\cos(1.5(\lambda-\phi)) $\\
		\hline
	\end{tabular}
\end{table}

\begin{table}[!t]
	\renewcommand{\arraystretch}{1.3}
	\caption{Sparsity Levels of Synthetic Maps (SMs)}
	\label{fig:SS_Sparsity}
	\centering
	\begin{tabular}{|c|c|c|c|c|c|}
		\hline
		 SM 1 &  SM 2 &  SM 3 &  SM 4 &  SM 5 \\
		\hline
		3 & 7 & 6 & 21 & 11 \\ \hline
	\end{tabular}
\end{table}
\subsection{TEC Map Generation Based on Compressive Sensing}
Although a TEC map (the matrix $\mathbf{U}$) is composed of $N=PQ$ pixels, assume that only $M$ of them are available as measured values, i.e. $M\leq N$. Let the pixels with measurements be represented by the set 
%\footnotesize 
\begin{equation}
\mathcal{O} = \{ (p_m,q_m) \;|\; p_m \in [0,P-1],\: q_m \in  [0,Q-1]\}
\end{equation}
%\normalsize
where $m=0,\cdots,M-1$ and $p_m,q_m \in \mathbb{Z}$. Let the measurements be stacked in an $M\times 1$ vector as 
\begin{eqnarray} \label{Eq:Observ}
\mathbf{b} &=& \begin{bmatrix}
b_0^{(i_0)}\ \\ \vdots \\b_m^{(i_m)} \\ \vdots \\ \quad b_{M-1}^{(i_{M-1})}
\end{bmatrix} =  \begin{bmatrix}
\mathbf{e}_{i_0}^T \mathbf{u} \\ \vdots \\ \mathbf{e}_{i_m}^T \mathbf{u} \\ \vdots \\\mathbf{e}_{i_{M-1}}^T \mathbf{u}
\end{bmatrix} \nonumber \\
&=&\begin{bmatrix}
\mathbf{e}_{i_0} \quad \mathbf{e}_{i_1} \quad \cdots \quad \mathbf{e}_{{i_M-1}}
\end{bmatrix}^T \mathbf{u} \nonumber \\
&=& \mathbf{M}\mathbf{u}
\end{eqnarray}
where $i_m = Pq_m+p_m$, and $\mathbf{e}_{i_m}$  denotes the $N\times 1$  vector with a 1 in the  $i_m$-th entry and $0$'s elsewhere, and $\mathbf{M}$ is the $M\times N$ measurement matrix. Substituting  (\ref{Eq:Transform}) into (\ref{Eq:Observ}) yields
\begin{equation}
\mathbf{b} = \mathbf{A} \mathbf{s}
\end{equation}
where the $M\times N$ matrix $\mathbf{A} = \mathbf{M} \mathbf{D} $ is called the sensing matrix.
%\begin{figure}[!t]
%	\centering
%%	\includegraphics[width=0.75\columnwidth]{./Figures/SS2_ERROR.eps}
%	\includegraphics[trim=0.5cm 0cm 1cm 0cm ,clip=true,width=\columnwidth]{./Figures/SS2_ERROR.eps}
%	\caption{Synthetic map 1: Normalized error energy}
%	\label{fig:SS2ERROR}
%\end{figure}

The conventional CS formulation\footnote{In a conventional CS formulation, the optimization problem would be $\underset{\mathbf{x}}{\text{arg min}} ||\mathbf{x}||_1$, subject to $||\mathbf{A}\mathbf{x}-\mathbf{b}||_2^2< \epsilon$, \cite{Candes}, where the cost function aims at minimizing the sparsity level of the problem in the transform domain (the actual formulation contains $||.||_0$ where $||.||_1$ is used as a mathematically tractable approximation of $||.||_0$, \cite{CS_TheAndApp}) and the constraint guarantees to limit the reconstruction error below a threshold level $\epsilon$.}, when used with the DCT basis, does not differentiate between low frequency and high frequency components. However, due to physical limitations, the TEC value cannot change abruptly in the spatial domain, i.e. a TEC map has a lowpass nature, as it is demonstrated in \cite{Yenen}.

\begin{figure}[!t]
	\centering
	\includegraphics[width=0.9\columnwidth]{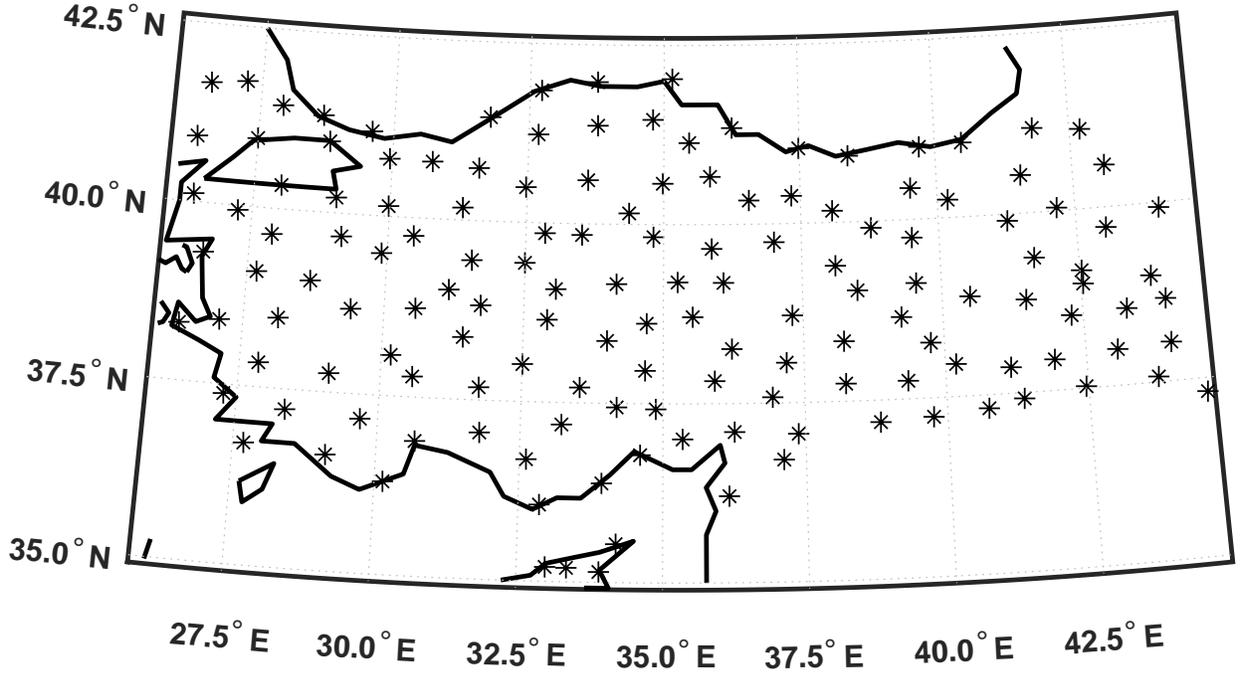}
	\caption{Reference station network of TNPGN-Active}
	\label{fig:ObservationPoint}
\end{figure}
To incorporate this lowpass characteristic to the problem, we propose a two part cost function as the following  
\begin{equation}
\begin{aligned}
%& \text{argmin}_\mathbf{x}
& \underset{\mathbf{s}}{\text{arg min}}
& & ||\mathbf{W} \mathbf{s}||_1 + \gamma f(\mathbf{D}\mathbf{s}) \\
& \text{subject to}
& &\frac{||\mathbf{A} \mathbf{s} -\mathbf{b}||_2^2}{||\mathbf{b}||_2^2}\leq \epsilon\\
%& & &\mathbf{D} \mathbf{x} = \mathbf{y}
\end{aligned}
\label{Eq:OptProb}
\end{equation}
where $\mathbf{W} = \text{diag}(1/w_0,1/w_1,\cdots,1/w_{N-1})$ is a square diagonal weighting matrix such that 
\begin{equation} \label{eq:wi_coeff}
w_{i} = \frac{1}{1+\frac{k^2+l^2}{\sigma^2}},\quad \text{for} \: i = Pl+k
\end{equation}
where the weights $w_i$ represent the coefficients of a first order Butterworth lowpass filter, and $\sigma$ is the cut-off wave number in both dimensions. Furthermore, we use a regularization function, $f(.)$, defined as
\begin{equation} \label{Eq:Regularization}
f(\mathbf{u}) = ||\nabla_x \mathbf{u} ||_2^2 + ||\nabla_y \mathbf{u} ||_2^2
\end{equation} 
to limit the variation in both spatial dimensions, where $\nabla_x$ and $\nabla_y$ are the gradient operators in x- and y-direction, respectively. We use CVX \cite{CVX1} as a solver for the problem in (\ref{Eq:OptProb}).
\vspace*{-0.2cm}
\section{Performance Evaluation of The Proposed Technique}
In this section, we will analyze the performance of the proposed CS based TEC map estimation technique given in (\ref{Eq:OptProb}). There are certain patterns of the ionosphere depending on both the time of the day, such as dawn, midday, dusk, and also the state of the ionosphere, such as those at a quiet day and at a disturbed day. Due to the time varying and random characteristic of the ionosphere, it might be difficult to observe these patterns explicitly. Therefore, we will first investigate the performance of our technique on synthetically generated maps which mimic the typical states of the ionosphere.
\vspace*{-0.4cm}
\subsection{Performance Evaluation over Synthetic Maps}
As a first set of performance evaluation, we have generated several synthetic TEC maps representing the typical patterns of the dawn (Fig. \ref{fig:SS2MAP}) and the midday (Figs. \ref{fig:SS3MAP} and \ref{fig:SS4MAP}) of a quiet day, and the midday (Fig. \ref{fig:SS6MAP}) and the dusk (Fig. \ref{fig:SS7MAP}) of a disturbed day. Expressions for generating the maps are given in Table \ref{fig:SS_Formulation}. Also, Table \ref{fig:SS_Sparsity} demonstrates the sparsity level of the TEC maps in Fig. \ref{fig:SS2MAP} to \ref{fig:SS7MAP}. Noting that there are $PQ = 1638$ pixels in each map, it is clear that the maps are highly sparse, justifying the suitability of CS for the reconstruction of TEC maps from few measurements. 
%\begin{figure}[!t]
%	\centering
%%	\includegraphics[width=0.75\columnwidth]{./Figures/SS6_ERROR.eps}
%	\includegraphics[trim=0.5cm 0cm 1cm 0cm ,clip=true,width=\columnwidth]{./Figures/SS6_ERROR.eps}
%	\caption{Synthetic map 4: Normalized error energy}
%	\label{fig:SS6ERROR}
%\end{figure}
%\begin{figure}[!t]
%	\centering
%%	\includegraphics[width=0.75\columnwidth]{./Figures/SS7_ERROR.eps}
%	\includegraphics[trim=0.5cm 0cm 1cm 0cm ,clip=true,width=\columnwidth]{./Figures/SS7_ERROR.eps}
%	\caption{Synthetic map 5: Normalized error energy}
%	\label{fig:SS7ERROR}
%\end{figure}
\begin{figure}[!t]
	\centering
	\includegraphics[width=0.9\columnwidth]{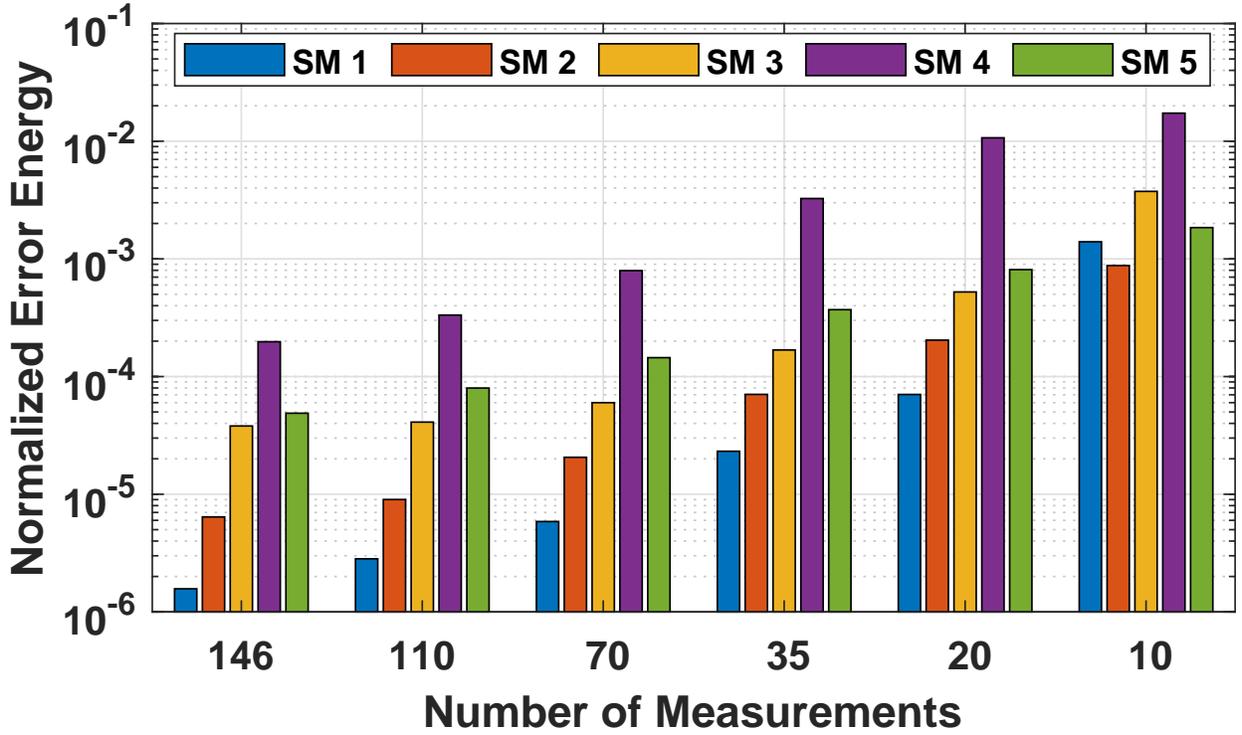}
	\caption{Effect of number of observation to the performance for each synthetic maps}
	\label{fig:SSObsernum}
\end{figure}

Fig. \ref{fig:ObservationPoint} shows the coordinates of the $146$ fixed GPS reference stations which constitute the TNPGN-Active CORS Network. We assess the performance of the proposed technique by taking samples from the synthetic maps at these coordinates only\footnote{We consider a resolution of $0.3^\circ$ (lat) $\times$ $0.3^\circ$ (lon). If the actual coordinates of a GPS receiver does not match a map grid point, we assign the measurement value to the nearest grid point. Although there may be other assignment methods, our approach is acceptable due to the low-pass characteristic of the TEC map and for a dense-enough grid structure}. Note the considerable amount of extrapolation done close to the boundaries of the map. The metric that we use for performance evaluation is the normalized error energy defined by
%\section{Synthetic Map Estimation Performance}
%In this section, synthetic maps given in previous section are used to evaluate the performance of compressive sensing algorithm. The observation points are taken from the Turkish National Permanent GPS Network (TNPGN). Reference station network given in Fig. \ref{fig:ObservationPoint} consists of $146$ continuously-operating Global Navigation Satellite System (GNSS) stations which are distributed nearly uniform across Turkey and North Cyprus Turkish Republic.
%
%Performance analysis of the optimization problem in (\ref{Eq:OptProb}) is measured with normalized error energy defined as
\begin{equation}
\text{\LARGE $\varepsilon$\normalsize$_{p}$} = \frac{||\mathbf{u} - \hat{\mathbf{u}}||_2^2}{||\mathbf{u}||_2^2}
\label{Eq:NormalizedError}
\end{equation}
where $\mathbf{u}$ is the synthetic map vector defined in (\ref{Eq:MapMatrix}) and (\ref{eq:u_vec}), and $\hat{\mathbf{u}}$ is the reconstructed map vector given by $\hat{\mathbf{u}} = \mathbf{D}\mathbf{s}$, and $\mathbf{s}$ is the solution of the problem in (\ref{Eq:OptProb}).

The performance of the algorithm depends on the values of the parameters $\sigma$ and $\gamma$ given in (\ref{Eq:OptProb}) and (\ref{eq:wi_coeff}). Examining over a wide range of values of  $\sigma$ and $\gamma$, 
%as can be seen in Figs \ref{fig:SS2ERROR} to \ref{fig:SS7ERROR},  
$(\sigma,\gamma) = (5,1)$ is found to be a good choice which yields relatively low reconstruction error for all synthetic maps. 
%Figs \ref{fig:SS2ERROR} to \ref{fig:SS7ERROR} reveal that the normalized error energy is below $7\times 10^{-4}$ for all conditions examined in Figs \ref{fig:SS2MAP} to \ref{fig:SS7MAP} and all $(\sigma,\lambda)$ values. For the optimized values $(\sigma,\lambda)=(5,1)$, the normalized error energy is at most $1.9\times10^{-4}$ which means a deviation of $0.019\%$ from the original map.
Our numerical analysis has revealed that this set results in a normalized error energy of $1.9\times10^{-4}$ as the worst case for SM4 and lower normalized error energy values for the other synthetic maps.

An important parameter that has effect on the performance of CS is the number of samples taken. Considering the number of GPS receivers in the TNPGN-Active network, we assume that at most $146$ samples (measurements) can be taken for a TEC map. However, due to several reasons in practice, only a subset of these receivers may be available. Fig. \ref{fig:SSObsernum} examines the performance of the proposed technique for various number of measurements. For each synthetic map scenario, and each number of measurements, $100$ different realizations are conducted with randomly chosen  samples from the full set of $146$ samples, and the normalized error energy metric is averaged over these realizations.

\begin{figure}[!t]
	\centering
	\includegraphics[width=0.85\columnwidth]{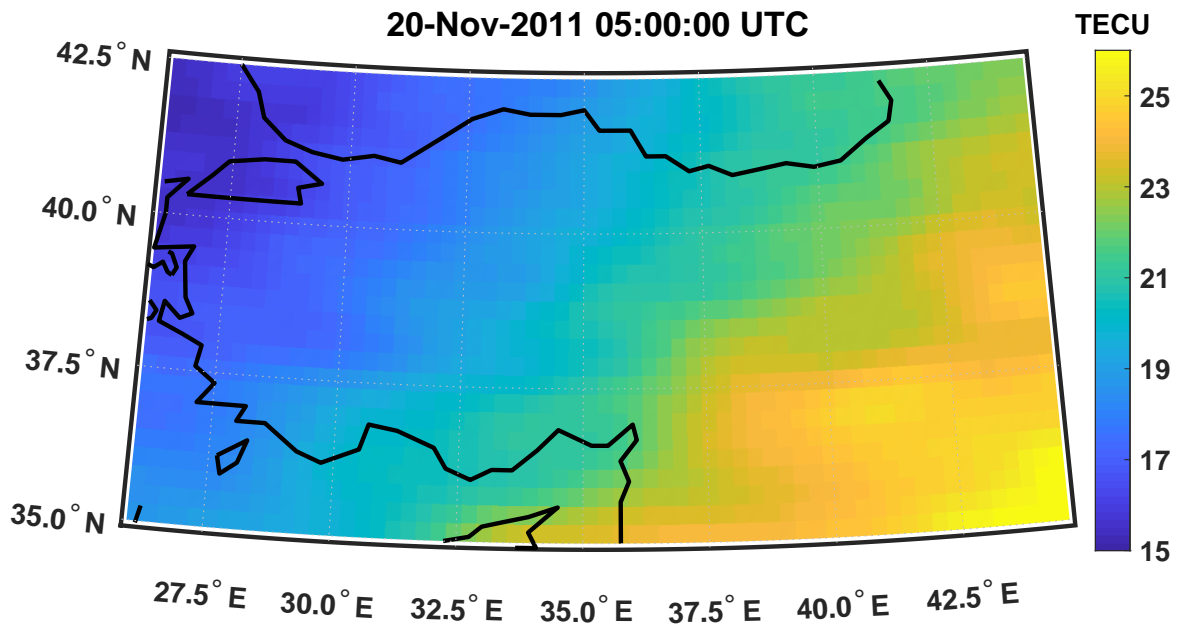}
	\caption{Estimated map at 05:00 UTC on 20 Nov 2011, Compressive Sensing }
	\label{fig:TECMAP_Day6_TIME0500UTC}
\end{figure}
\begin{figure}[!t]
	\centering
	\includegraphics[width=0.85\columnwidth]{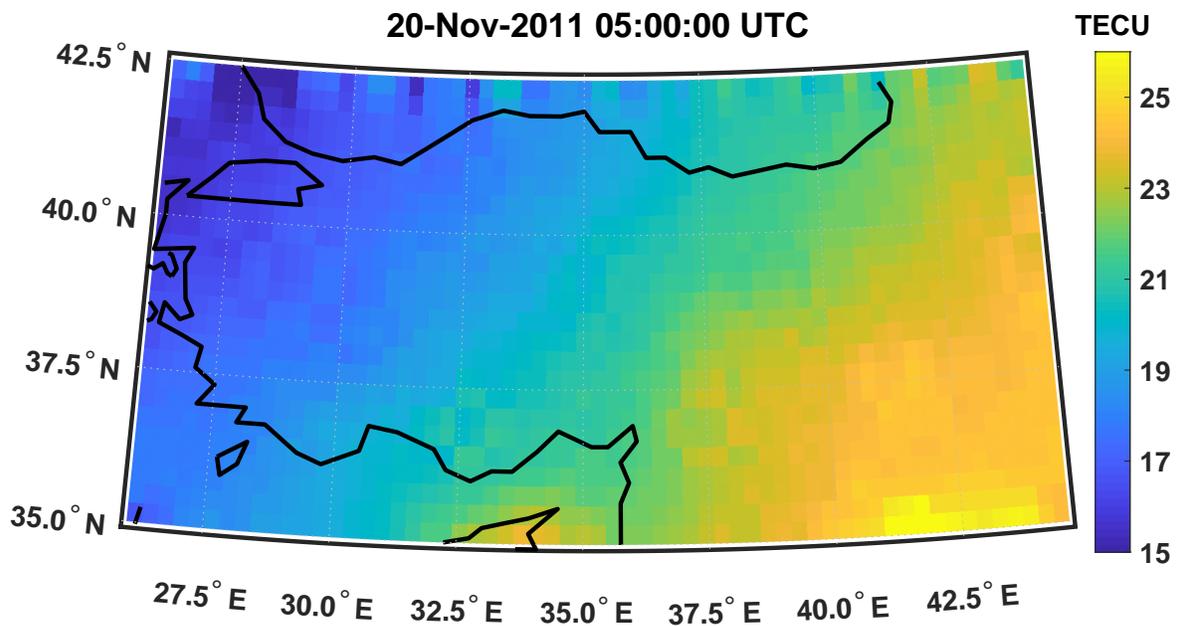}
	\caption{Estimated map at 05:00 UTC on 20 Nov 2011, Kriging}
	\label{fig:TECMAP_Day6_TIME0500UTC_Krig}
\end{figure}
Fig. \ref{fig:SSObsernum} demonstrates that even if half of the samples (i.e. $70$ samples) are available, the normalized error energy is below $4\times10^{-4}$ for all synthetic maps. Considering the sparsity levels in Table \ref{fig:SS_Sparsity}, this is coherent with the general observation that the number of samples approximately equal to five times the sparsity level is adequate for a good reconstruction \cite{CS_TheAndApp,Candes,Rani}.
\vspace*{-0.2cm}
\subsection{Performance Evaluation over Measurement Data}
By using the RINEX measurements of the GPS receivers at the TNPGN-Active network (Fig. \ref{fig:ObservationPoint}), we first calculate the TEC estimates at the sampling points in the local zenith direction by using the IONOLAB-TEC algorithm \cite{GPS1,GPS2,TNGPN2,TNGPN3}, for 25 June 2011 16:40 UTC and 20 November 2011 05:00 UTC. IONOLAB-TEC is a state-of-the-art signal processing technique  that calculates the TEC value in the local zenith direction at a specific location (e.g. a CORS GPS receiver) by combining IONOLAB-BIAS and Vertical TEC (VTEC) for each satellite in the least square sense by  employing a weighting function. VTEC is the total number of electrons in the local zenith direction obtained from Slant TEC (STEC) by the use of a "mapping function". The number of available measurements are respectively 122 and 124 out of the 146 sampling points for the epochs mentioned above.
\begin{figure}[!t]
	\centering
	\includegraphics[width=0.85\columnwidth]{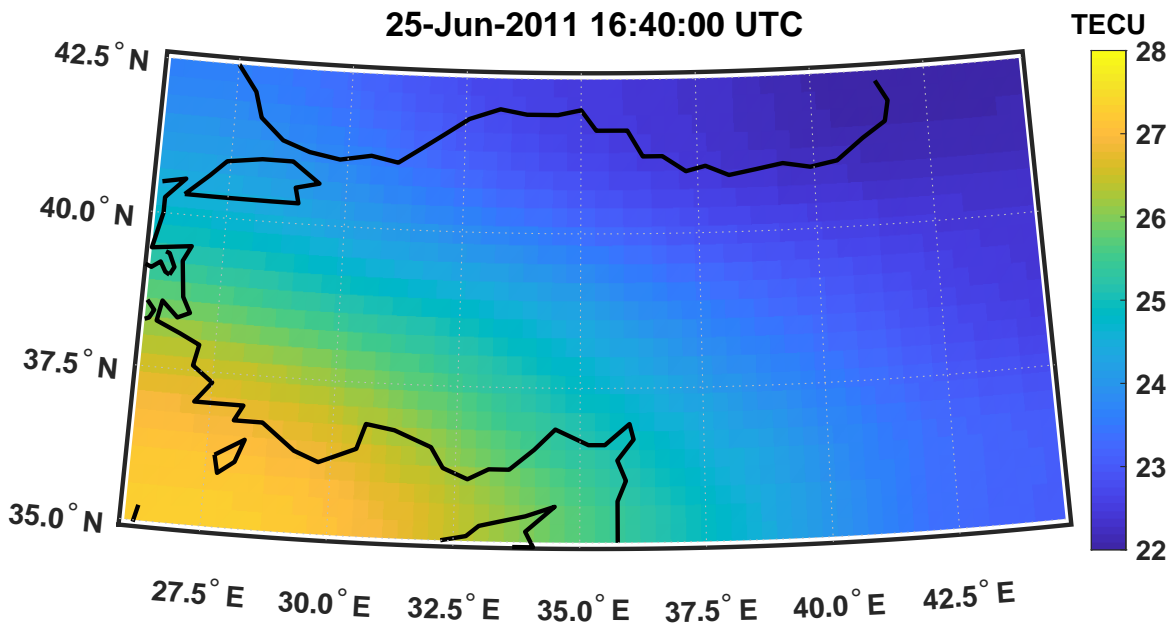}
	\caption{Estimated map at 16:40 UTC on 25 Jun 2011, Compressive Sensing }
	\label{fig:TECMAP_Day2_TIME1640UTC}
\end{figure}
\begin{figure}[!t]
	\centering
	\includegraphics[width=0.85\columnwidth]{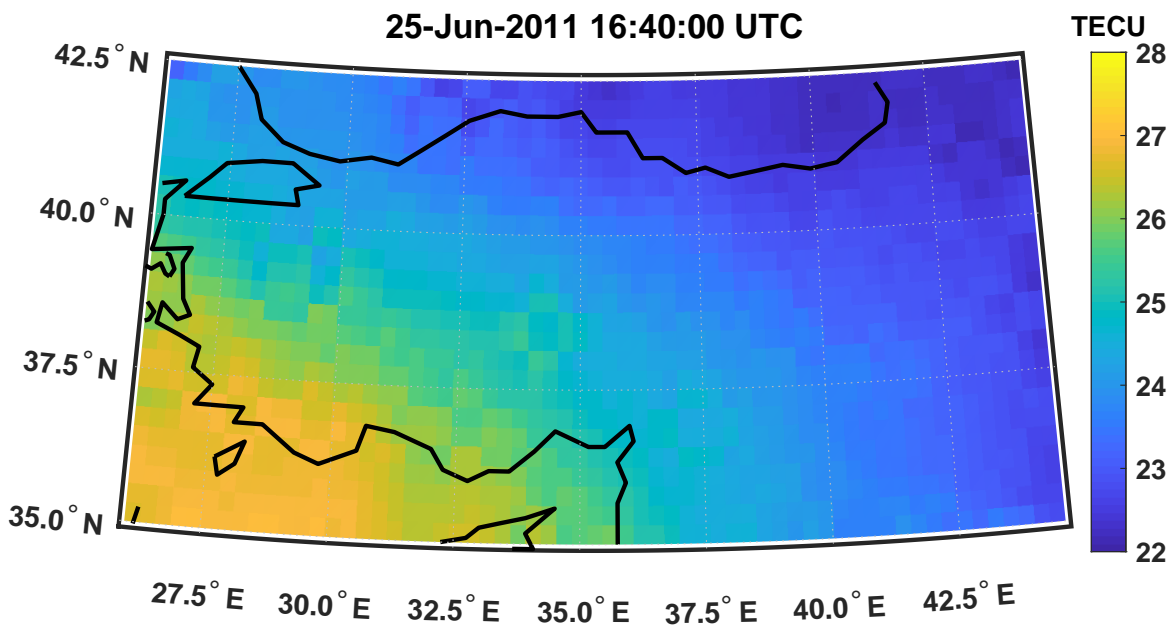}
	\caption{Estimated map at 16:40 UTC on 25 Jun 2011, Kriging}
	\label{fig:TECMAP_Day2_TIME1640UTC_Krig}
\end{figure}

TEC maps generated by the proposed CS based technique are given in Figs. \ref{fig:TECMAP_Day6_TIME0500UTC} and \ref{fig:TECMAP_Day2_TIME1640UTC}. We evaluate the performance of the proposed technique in two ways.

\begin{figure}[!t]
	\centering
	\includegraphics[width=0.8\columnwidth]{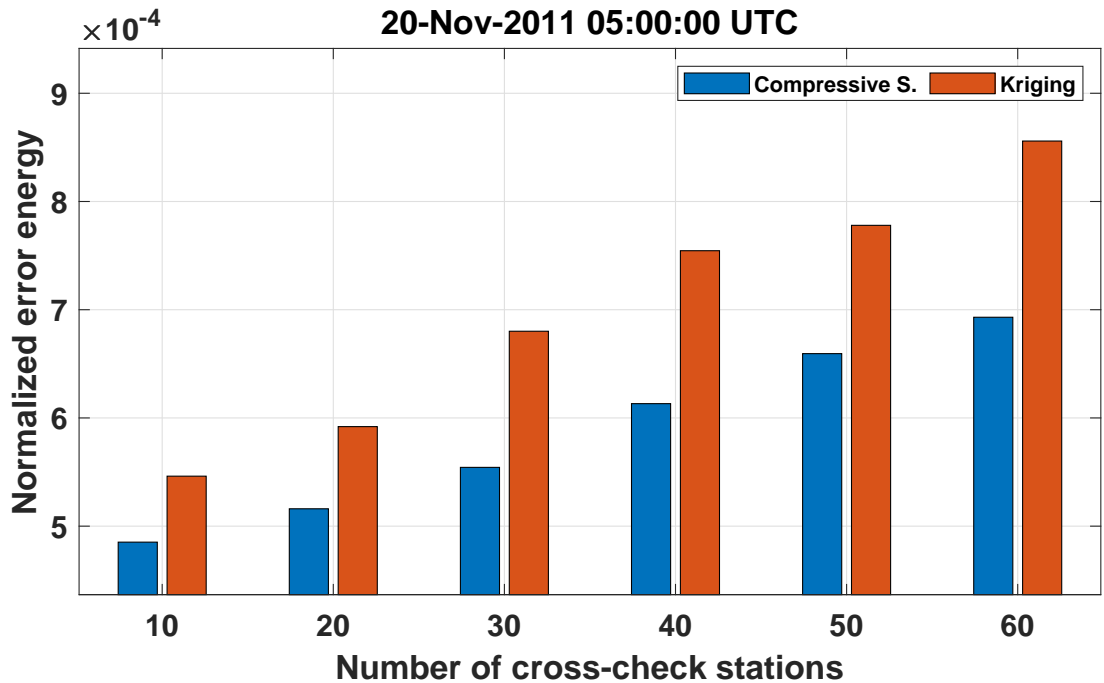}
	\caption{Cross check at 05:00 UTC on 20 Nov 2011}
	\label{fig:TECMAP_Day6_TIME0500UTC_CC}
\end{figure}

\begin{figure}[!t]
	\centering
	\includegraphics[width=0.8\columnwidth]{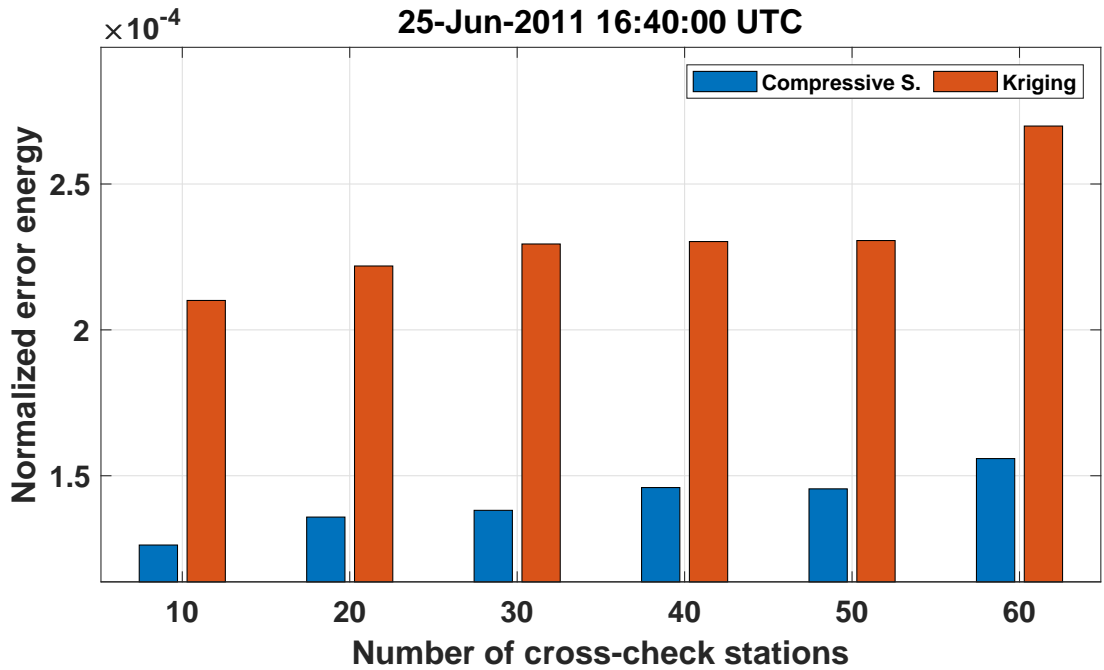}
	\caption{Cross check at 16:40 UTC on 25 Jun 2011}
	\label{fig:TECMAP_Day2_TIME1640UTC_CC}
\end{figure}

First, a number of measurements are randomly selected and kept out of the dataset that we use for generating the TEC maps with the proposed algorithm. Next, these measurements are compared with the estimates at the same coordinates of the TEC map in terms of the normalized error energy given in (\ref{Eq:NormalizedError}). We repeat this procedure 1000 times and take the average of the normalized error energy. The results are provided in Figs. \ref{fig:TECMAP_Day6_TIME0500UTC_CC} and \ref{fig:TECMAP_Day2_TIME1640UTC_CC} for different number of cross-check points. It can be seen that with 10 to 30 cross-check points, the error energy is below $1.85\times 10^{-4}$ for the epoch in Fig. \ref{fig:TECMAP_Day2_TIME1640UTC_CC} and below $5.6\times 10^{-4}$ for the epoch in Fig. \ref{fig:TECMAP_Day6_TIME0500UTC_CC} which are reasonably low, justifying the excellent performance of the proposed algorithm. Note that, as we increase the number of cross-check points, the performance deteriorates since the number of measurements included in the dataset of the algorithm decreases as it was demonstrated in Fig. \ref{fig:SSObsernum} for the synthetic maps.


Kriging, interpolation with spherical harmonics, thin plate spline interpolation and neural networks \cite{Orus2005,Wielgosz2003,Schaer1999,Sayin2008,MGurun2007,Deviren2018} are some of the methods used for TEC map generation in the related literature. For a second evaluation, we compare the performance of the proposed technique with the state-of-the-art technique based on Kriging \cite{MGurun2007,Deviren2018}. Kriging algorithm uses inverse distance wieghting to increase the number of samples in the dataset then generates TEC map by using ordinary Kriging with isotropic semivariogram. TEC maps generated by Kriging with the same dataset that we use to generate the TEC maps with our proposed technique, are shown in Figs. \ref{fig:TECMAP_Day6_TIME0500UTC_Krig} and \ref{fig:TECMAP_Day2_TIME1640UTC_Krig}. Artificial fluctuations can be clearly seen by comparing Figs. \ref{fig:TECMAP_Day6_TIME0500UTC_Krig}-\ref{fig:TECMAP_Day2_TIME1640UTC_Krig} and \ref{fig:TECMAP_Day6_TIME0500UTC}-\ref{fig:TECMAP_Day2_TIME1640UTC}, especially for the regions close to the edges of the map. Such fluctuations are not physically possible due to the nature of the ionosphere.

We also conducted a cross-check test similar to that is done for the proposed CS technique. Cross-check results for Kriging are also provided in Figs. \ref{fig:TECMAP_Day6_TIME0500UTC_CC} and \ref{fig:TECMAP_Day2_TIME1640UTC_CC}. It can be concluded that the normalized error energy performance of our technique is better than that of Kriging, justifying the superiority of our technique.

\vspace*{-0.6cm}
\section{Conclusion}
In this paper, a modified compressive sensing technique is proposed for generating regional TEC maps by using a dataset obtained from a CORS network. The algorithm is based on the observation that TEC maps have a lowpass structure, and they can be highly sparse in the 2D-DCT domain. 

To examine the performance of the proposed technique, we use synthetically generated TEC maps which mimic some of the common characteristics of the ionosphere. Several parameters of the technique, such as the weights of the algorithm, number of measurements required, are tuned over these maps. Analysis has revealed that the normalized-error-energy can be as low as $10^{-5}$ to $10^{-4}$.

We further use actual TNPGN-Active CORS network measurement data to assess the performance of the proposed technique. A cross-check based method is adopted and it is shown that the proposed technique can estimate the TEC map with high fidelity, even for a very high resolution of $0.5^\circ$ (lat) $\times$ $0.5^\circ$ (lon). We also compare the CS based technique with the state-of-the-art in the literature, a Kriging based algorithm. The proposed technique outperforms the Kriging based algorithm both in terms of normalized-error-energy and yields better perception and also fits better to the physical behaviour of the ionosphere.

Our analysis with measurement data also revealed that approximately 100 GPS-CORS measurements are adequate to map a region of $7.5^\circ$ (lat) $\times$ $20^\circ$ (lon) at a resolution of $0.5^\circ$ (lat) $\times$ $0.5^\circ$ (lon), i.e. 1638 pixels.

%Since the kriging is an estimator of geostatistics and majorly used in TEC map estimations, the CS estimates are compared with the kriging method. Kriging results in spurious squared patterns. These patterns are eliminated in the CS estimates and more smooth varying maps are estimated. 

% if have a single appendix:
%\appendix[Proof of the Zonklar Equations]
% or
%\appendix  % for no appendix heading
% do not use \section anymore after \appendix, only \section*
% is possibly needed

% use appendices with more than one appendix
% then use \section to start each appendix
% you must declare a \section before using any
% \subsection or using \label (\appendices by itself
% starts a section numbered zero.)
%

%\appendices
%\section{Proof of the First Zonklar Equation}
%Appendix one text goes here.
%
%% you can choose not to have a title for an appendix
%% if you want by leaving the argument blank
%\section{}
%Appendix two text goes here.
%
%
%% use section* for acknowledgment
%\section*{Acknowledgment}
%
%
%The authors would like to thank...

% Can use something like this to put references on a page
% by themselves when using endfloat and the captionsoff option.
\ifCLASSOPTIONcaptionsoff
  \newpage
\fi

% trigger a \newpage just before the given reference
% number - used to balance the columns on the last page
% adjust value as needed - may need to be readjusted if
% the document is modified later
%\IEEEtriggeratref{8}
% The "triggered" command can be changed if desired:
%\IEEEtriggercmd{\enlargethispage{-5in}}

% references section

% can use a bibliography generated by BibTeX as a .bbl file
% BibTeX documentation can be easily obtained at:
% http://mirror.ctan.org/biblio/bibtex/contrib/doc/
% The IEEEtran BibTeX style support page is at:
% http://www.michaelshell.org/tex/ieeetran/bibtex/
\bibliographystyle{IEEEtran}
\end{document}